# Capacity Fade due to Side-reactions in Silicon Anodes in Lithium-ion Batteries

*Vijay A. Sethuraman*[1,*]

*Environmental Energy Technologies Division*
*Lawrence Berkeley National Laboratory*
*1 Cyclotron Road, Berkeley, CA 94720-8168, USA*

It is shown that continuously occurring electrolyte-reduction reaction on freshly-exposed electrode surfaces during lithiation/delithiation cycles causes the lowering of cycling efficiency, and hence, capacity fade in *well-cycled* silicon anodes in lithium-ion batteries. Using galvanostatic lithiation/delithiation data from multiple cycles on a Li/Si half-cell, a methodology to separate the charge due to the main reaction (lithiation/delithiation of Si) from the side-reaction (electrolyte-reduction) is presented. The rate of this parasitic side reaction is estimated on *well-cycled* amorphous silicon thin-film electrodes at ambient temperature for the following three commonly-used lithium-ion electrolyte formulations: mixtures of ethylene carbonate and diethylene carbonate (EC:DEC) with and without a fluoroethylene carbonate (FEC) additive, and propylene carbonate (PC), all containing 1.2 M lithium hexafluorophosphate. Among the three formulations, the electrolyte containing EC:DEC with the FEC additive exhibits the lowest coulombic losses due to side-reactions, followed by PC, and EC:DEC without the FEC additive (*i.e.,* EC:DEC + FEC > PC > EC:DEC). The importance of estimating side-reaction rates on a *well-cycled* electrode is discussed in the context of self-discharge, capacity fade, development of battery management system algorithms and precise mathematical modeling of lithium-ion batteries.

[1]Present address: Box D, School of Engineering, Brown University, 182 Hope Street, Providence, Rhode Island 02912, USA.
[*] *E-mail* – vj@cal.berkeley.edu (V.A. Sethuraman); Tel. +1 510 764 4842

## 1. Introduction

Silicon's maximum lithiation capacity of 3579 mAh $g^{-1}$ [1,2] is much higher than that of graphite (372 mAh $g^{-1}$) [3,4], making it an attractive choice for use as negative electrodes in lithium-ion batteries. A large body of literature exists on the electrochemical and mechanical performance of anodes made of pure silicon and composites in which silicon is one of the constituents [ 5 ]. The electrochemical lithiation and delithiation of silicon at ambient temperatures have been extensively studied in the recent years in such forms as nanowires [6-8], amorphous thin films [9-11], crystalline thin films [12] crystalline powder [13,14], composites [15-17], mixtures with other metals [18-20], and mixtures with carbon [21-23].

Regardless of the nature of silicon geometry studied, the following common features appear in all of the studies on its electrochemical lithiation/delithiation: (a) high first-cycle irreversible loss in capacity (*ca.* 40-60%); (b) delithiation potential being higher than lithiation potential (*vs.* $Li/Li^+$) at any state-of-charge (SOC); (c) huge volume expansion during lithiation (370% when fully charged); and (d) gradual loss of available active material upon cycling (*i.e.,* cycle-to-cycle loss of capacity) once the cell had reached a steady cycling efficiency. Furthermore, silicon electrodes have been recently shown to undergo cycles of compressive and tensile stresses with large plastic strains [24,25,26,27,28,29,30], the mechanics of which plays a huge role in the above mentioned characteristics. The large volume expansion during lithiation and delithiation of Si causes repetitive straining of the thin solid-electrolyte-interphase (SEI) layers that form during the first few cycles. Unlike graphite electrodes where the SEI layer formed during the first cycle remains stable, the SEI layer formed on Si gets damaged because of large strains and continues to re-form during subsequent cycles. This results in the consumption of lithium ions and hence capacity fade.

Obrovac and Krause [13] have shown that the capacity loss in silicon anodes could be minimized by adopting proper lithiation and delithiation cut-off potentials during cycling. This is because amorphous $Li_xSi$ crystallizes very rapidly to a new $Li_{3.75}Si$ phase below 50 mV *vs.* $Li/Li^+$ [1,2] and, although this crystalline phase converts back to amorphous $Li_xSi$ during subsequent delithiation, some irreversibility exists. Besides, adapting proper cut-off potentials can also limit volume expansion of Si [13]. The difference between the lithiation and delithiation capacities is greatest for the first cycle, which can be attributed to the formation of the SEI layer (*i.e.*, side reactions involving the reduction of the electrolyte below 0.8 – 1.0 V *vs.* $Li/Li^+$), and loss due to accumulation of $Li^+$ into the bulk of the material electronically isolated due to electrode delamination and/or pulverization.

However, on a *well-cycled* silicon electrode with no variation in cycle-to-cycle discharge capacity (*i.e.,* after the first few cycles where irreversible losses due to pulverization and SEI formation on a much larger surface area accounts for greater irreversible losses), the total lithiation capacity of the electrode between any two arbitrarily chosen cutoff potentials between (1.2 V and 0 V *vs.* $Li/Li^+$ in this study) is always higher than the total delithiation capacity between the same two cutoff potentials. A direct consequence of this is that the efficiency of the lithiation/delithiation cycle is always lower than 100%. Though this is true for all negative electrodes in lithium-ion batteries, full-cell cycling efficiency of state-of-the-art lithium-ion batteries with graphite anodes is upwards of 99.99% [31], while that of the best silicon anodes is only 99.95 [32]. Though the difference between these two efficiency numbers appears to be insignificant, the respective capacity retention at the end of 500 cycles is 95% and 77.8%. This is because capacity retention of a lithium-ion battery is proportional to $\eta^N$, where N represents the total number of cycles; and $\eta$, the cycling efficiency. Any departure of cycling efficiency from 100% represents the loss of cycleable lithium-ions, which results in capacity fade. In view of this, the origin of this decreased cycling efficiency should be determined, and its cause quantified.

It is shown in this study that electrolyte reduction on freshly exposed surfaces, accounts for the lowering of cycling efficiency once the cell has reached a steady cycling, where cycle-to-cycle capacity remains constant. The rate of this side-reaction on silicon anodes is estimated for three different electrolyte formulations commonly used in the lithium-ion-battery industry, and the activation energy for this side reaction is estimated for the formulation that exhibits the best cycling efficiency. The practical importance of estimating side-reaction rates is discussed in the context of self-discharge, capacity fade, development of battery management system algorithms and precise mathematical modeling of lithium-ion batteries.

## 2. Experimental

### *2.1. Electrode fabrication*

Thin copper discs (15.875 mm diameter, 0.3048 mm thick) were used as substrates for electrode fabrication. Silicon thin films were prepared by RF-magnetron sputtering (Edwards Auto 306 Sputter Coater) of a silicon target (76.2 mm diameter, 12.7 mm thick disc, 99.995% Si, Plasmaterials Inc., Livermore, California) at 200 W power, and at a pressure of 0.667 Pa of Argon (99.995%). Copper thin films were prepared by DC-sputtering of copper target (76.2 mm diameter, 3.175 mm thick disc, 99.995%, Super Conductor Materials Inc., Suffern, New York) at 100 W, and a pressure of 0.013 Pa of Argon. A *ca.* 300 nm copper thin film (*i.e.,* a Cu underlayer) was first sputtered onto the copper disc followed by the deposition of *ca.* 500 nm silicon film. Previous studies show that the Cu underlayer is critical to the continuous cycling of Si thin films [33,34]. The thicknesses of the sputtered films were monitored continuously during deposition *via* resonant-frequency changes of a gold crystal placed inside the sputtering chamber.

### *2.2. Electrode characterization*

The magnetron-sputtered Si thin-films were examined in a high-resolution JEOL JSM-6340F field-emission scanning electron microscope operated at an accelerating voltage of 15 kV using 5 mm as the working distance with the secondary and backscattered electron-image detectors. Energy dispersive X-ray (EDX) spectroscopy was carried out using a Genesis XM2 microanalysis system (EDAX Inc., Mahwah, NJ) to evaluate the surface composition of the film. The surface was also analyzed by Raman microscopy (Labram, ISA Groupe Horiba) with a helium-neon (HeNe) laser ($\lambda$ = 632.8 nm) at 1 mW power as the excitation source. The thickness of the magnetron-sputtered Si thin-films was determined from cross-sectional SEM images. In order to obtain a clean cross-sectional splices without damaging the Si film, a protective layer of Cu was sputtered on the Si film.

### *2.3. Coin-cell assembly*

To ensure the complete removal of residual moisture, the sputtered discs were baked at 393.15 K for 24 hours or more in vacuum. A sputtered-disc was then transferred into a glove-compartment without exposing it to air, and assembled into a 2032 coin cell configuration (*i.e.,* 20 mm diameter and 3.2 mm total thickness, National Research Council, Canada) under Argon atmosphere with a lithium-metal counter and reference electrode, and a woven Celgard 2500 separator (Celgard Inc., Charlotte, North Carolina). 1.2 M lithium hexafluorophosphate dissolved in one of the following three solvents was used as the electrolyte: (a) mixture (1:2 by % wt.) of ethylene carbonate and diethyl carbonate (EC:DEC), (b) propylene carbonate (PC), and (c) mixture (3:7 by % wt.) of ethylene carbonate and diethyl carbonate with 10% fluoro-ethylene carbonate ((EC:DEC). The FEC additive had been shown to increases the cycling efficiency of silicon anodes, probably due to the formation of a stable solid-electrolyte-interphase (SEI) layer [34,35,36]. All electrolyte formulations were obtained pre-mixed from Novolyte Technologies (Independence, Ohio).

*2.4. Galvanostatic cycling*

Electrochemical measurements were conducted at 23°C (±1°C) using a Solartron 1480A MultiStat system (Solartron Analytical, Oak Ridge, Tennessee), and data acquisition was done using Corrware (Scribner Associates Inc., Southern Pines, North Carolina). The cell was cycled galvanostatically at 25 $\mu A/cm^2$ (geometric area, *ca.* C/8 rate; C/8 rate corresponds to a current allowing a full discharge in 8 hours) total current between 0.01 and 1.2 V *vs.* Li. A lower cut-off potential of 10 mV *vs.* $Li/Li^+$ was chosen to prevent possible lithium deposition and to avoid the formation of large amounts of the crystalline $Li_{15}Si_4$ phase. Data acquisition rate was 1 Hz for all the electrochemical experiments. Galvanostatic cycling as described above was repeated between 5 °C and 45 °C in ten-degree intervals for estimating the activation energy.

## 3. Results and discussion

Characterization studies (elemental composition, conductivity, crystallinity, surface morphology, thickness, *etc.*) of the magnetron-sputtered silicon thin-film electrodes have been reported previously [34], and will only be discussed minimally here.

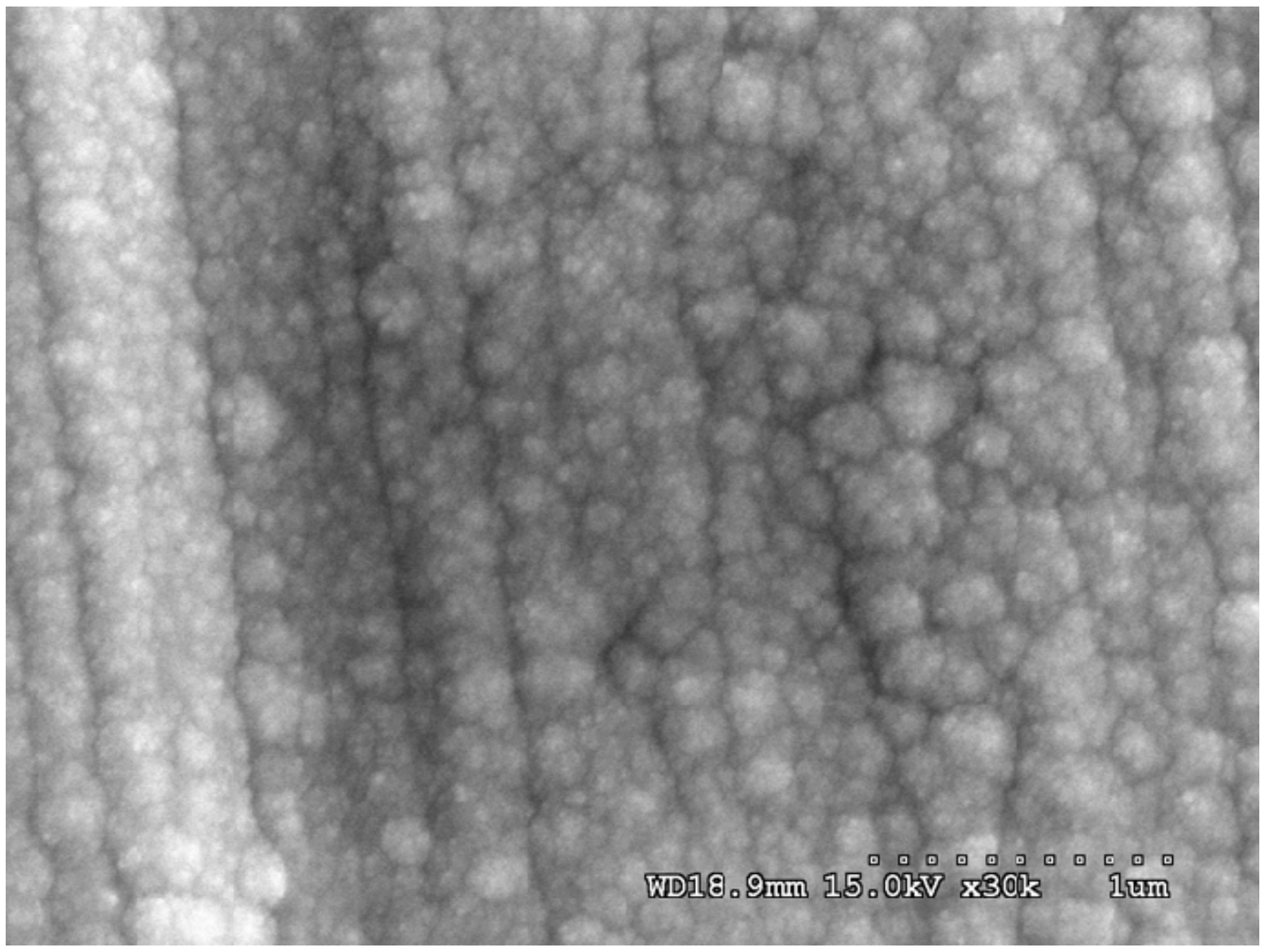

*Figure 1: Scanning-electron micrograph of magnetron-sputtered Si thin film. The length of the scale bar is 1 μm.*

Figure 1 shows the scanning electron micrograph of the magnetron-sputtered Si thin film. The film is made of nano-crystalline grains [see figure 2 in reference 34] with a size distribution in the range of *ca.* 50-300 nm. Raman spectrum (not shown) does not show a sharp peak corresponding to 520 $cm^{-1}$ indicating that the film does not have a long-range order. Though the film appears to be porous, no attempt was made to characterize the porosity of the film and the applied current-density and the rate constant for the side reaction ($i_{0,side}$) were normalized to the geometric area of the thin-film electrode. The thickness of the sputtered Si thin films was measured using cross-sectional SEM images of the thin-film electrode structure, a sample of which is shown in Figure 2. A protective top Cu layer was sputtered on the Si film prior to making splices for cross-sectional imaging.

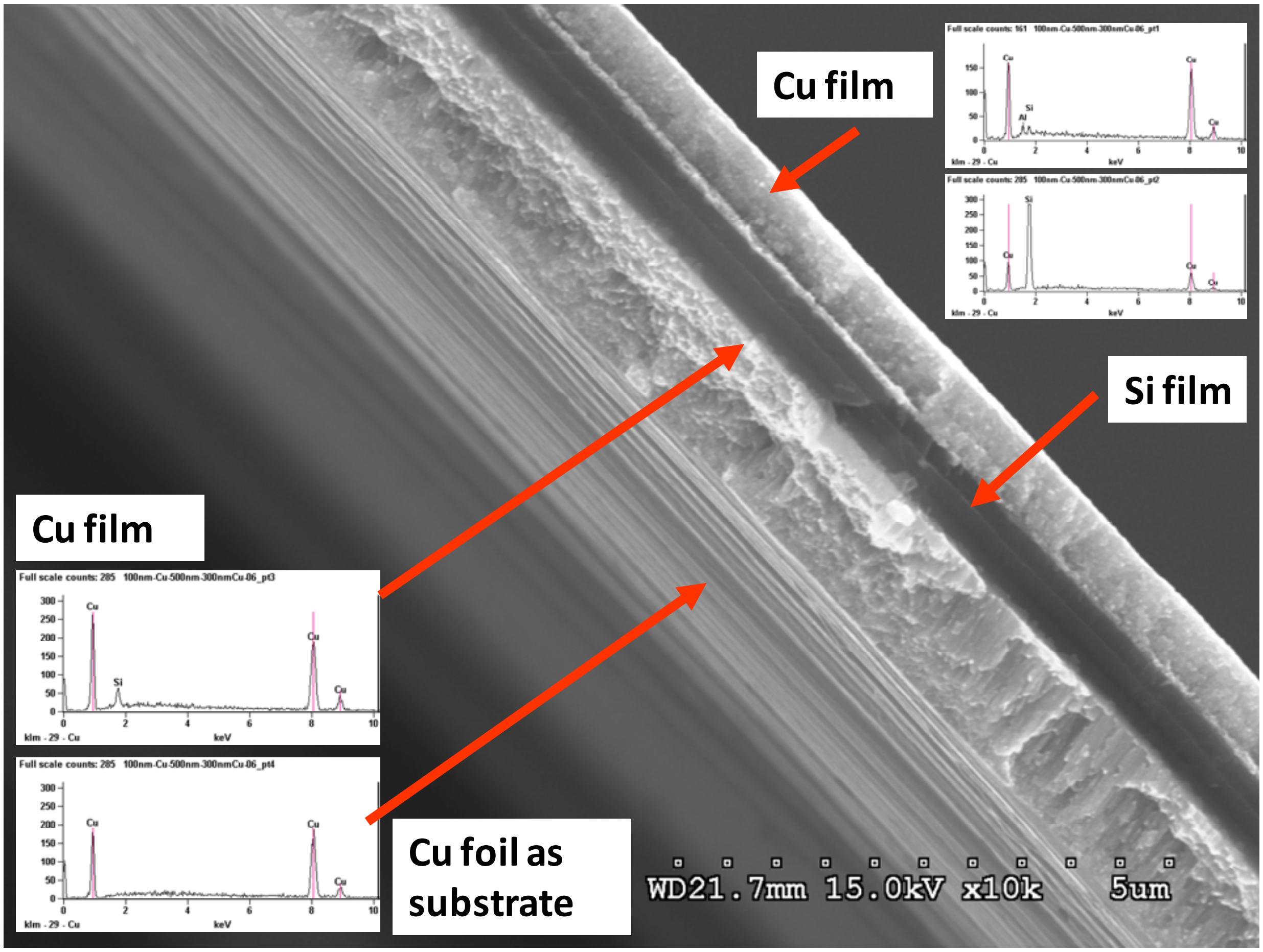

*Figure 2: Cross-sectional scanning-electron micrograph of the magnetron-sputtered Si thin-film-electrode structure. The inset graphs display the energy-dispersive X-ray spectra of the respective layers. The units on the ordinate of the inset graphs are arbitrary and the units on the abscissa are in keV. Note that the top copper layer was present only for the purposes of making a cross-sectional scanning-electron micrograph and not during the electrochemical-cycling studies. The thickness of the Si film as shown is ca. 625 nm.*

The thicknesses were verified using step profilometer (AlphaStep D-100 Stylus Profiler) measurements. Thicknesses from both the cross-sectional SEM images and the profilometer measurements correlated well with the thickness estimates from the crystal monitor inside the deposition chamber. The capacity of the Si thin-film electrode was calculated using the measured thickness and assuming bulk properties for the film (*i.e.*, density of 2.33 g/cm$^3$).

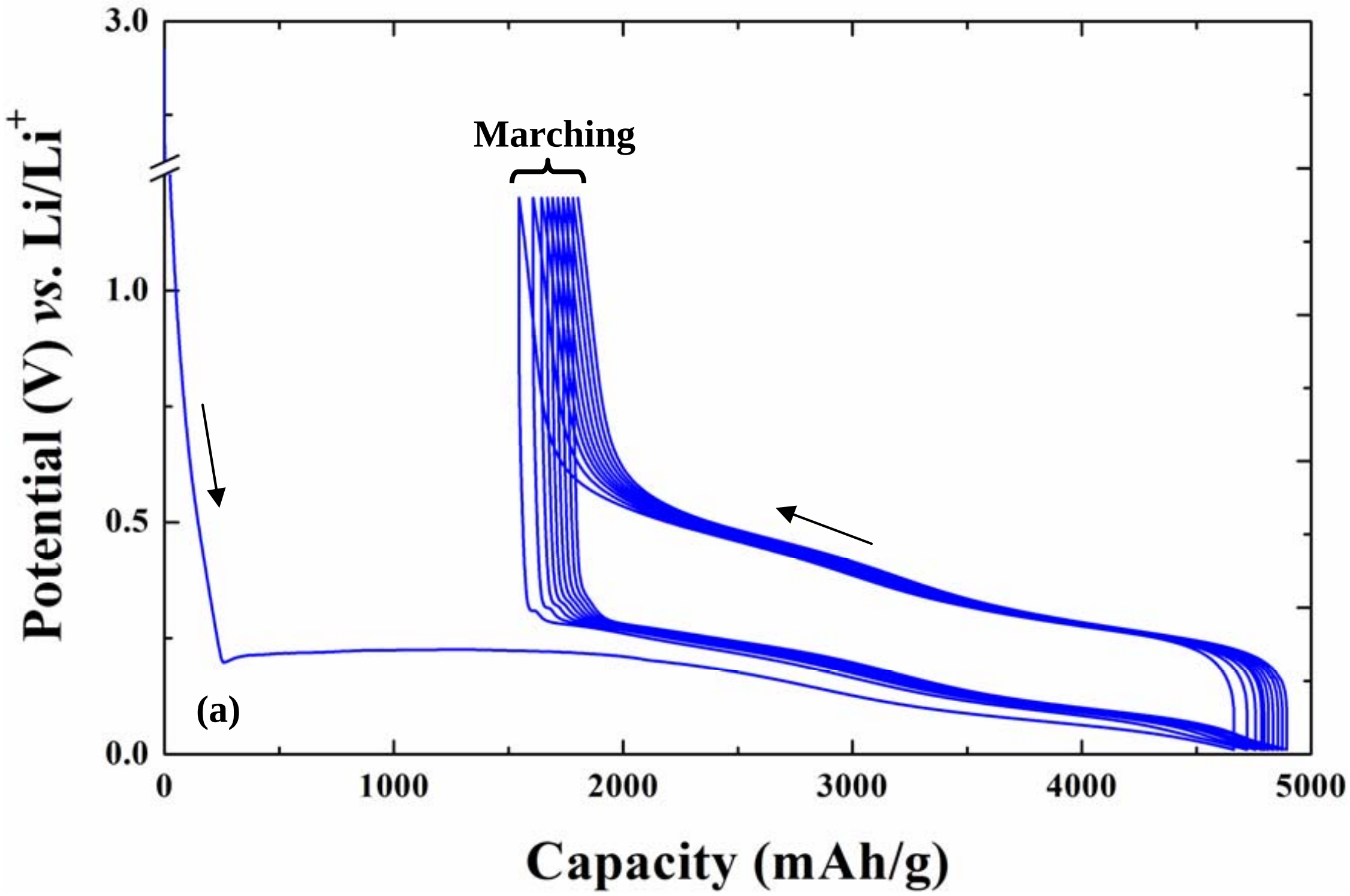

*Figure 3: (a) Cell potential vs. capacity curves corresponding to lithiation and delithiation of a magnetron-sputtered, silicon thin-film electrode cycled at C/8 rate between 1.2 and 0.01 V vs. Li/Li$^+$. Data corresponding to the first ten cycles are shown. The electrolyte used in this case is 1.2 M LiPF$_6$ in EC:DEC (1:2) with 10% FEC. The arrows indicate the cycling direction.*

Figure 3a shows potential *versus* capacity curves of the sputtered silicon thin-film electrode cycled at *ca.* C/8 rate between 1.2 and 0.01 V *vs.* Li/Li$^+$. The electrolyte in this case is 1.2 M LiPF$_6$ in EC:DEC with 10% FEC. The first-cycle irreversible capacity loss is approximately 35%, which is lower than that of composite electrodes made with crystalline silicon powder [13]. The cycling behavior of the sputtered Si thin film from the second cycle onwards is very similar to that of composite electrodes, albeit with a lower potential hysteresis (*ca.* 250 mV *vs.* 320 mV for the composite electrodes [13], both at 50% SOC). The crystallization of Li$_x$Si as seen by Obrovac and Krause [13] and the associated irreversible loss of active Si do not appear to take place here.

The lithiation and delithiation capacities, as well as the cycling efficiency numbers are summarized in Figure 3b. Similar to the electrochemical performance of various forms of amorphous silicon electrodes reported in the literature, the magnetron-sputtered silicon thin-film electrode exhibits larger reversible capacity (*ca.* 3100 mAh/g) from second cycle onwards; the cycling efficiency goes up above 99% for the subsequent cycles. Form cycle 7 onwards, both lithiation and delithiation capacities remain constant. However, the delithiation capacity is consistently lower than lithiation capacity by *ca.* 24 mAh/g (or 0.8%) for each of those cycles. This is reflected in the cycle-to-cycle marching behavior seen in Figure 3a.

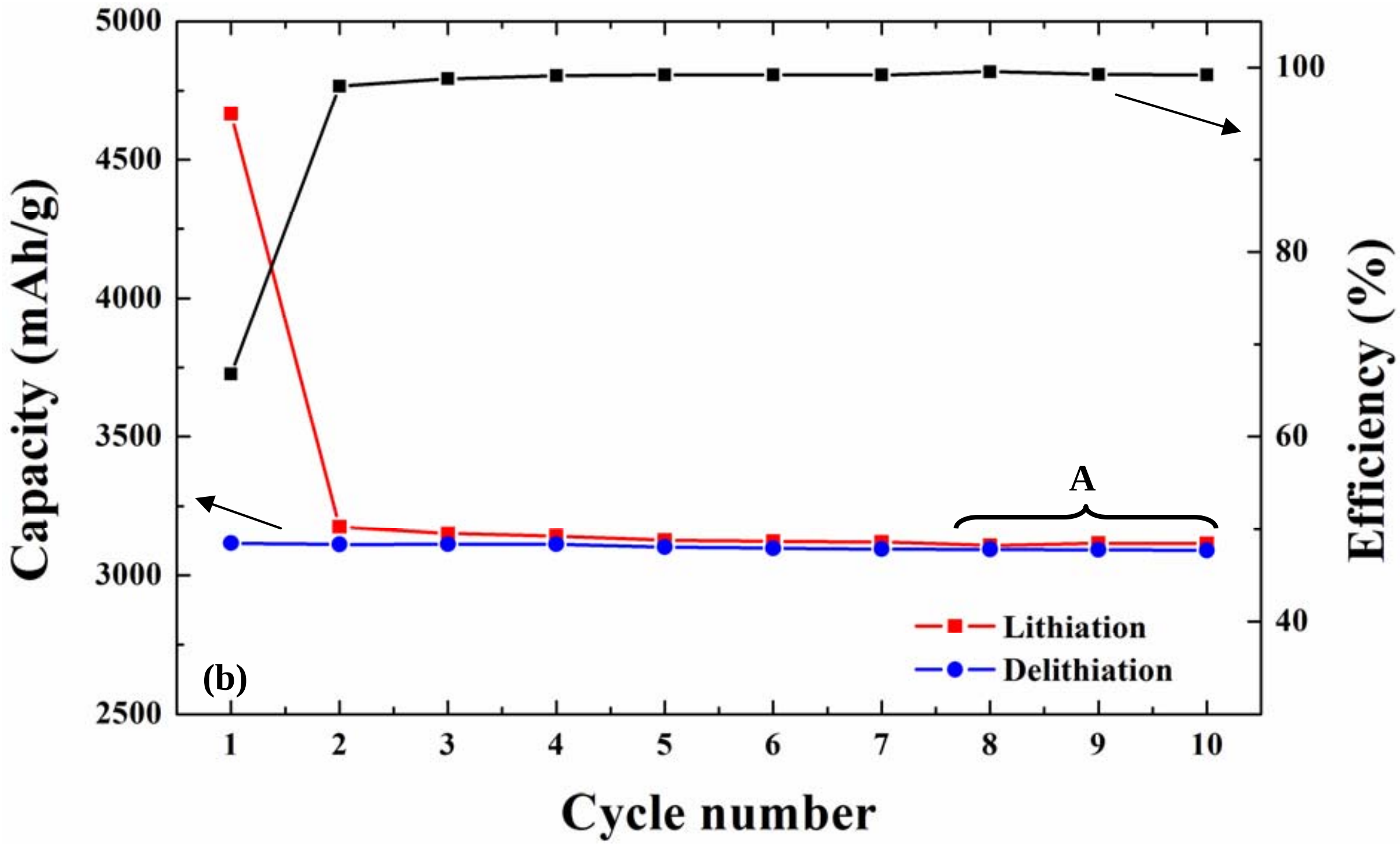

*Figure 3(b): Lithiation and delithiation capacities (left axis) for the first ten cycles are shown along with the respective cycling efficiency (right axis). These numbers correspond to data shown in figure 3a. Data marked A represents the region of no cycle-to-cycle capacity variation where the side-reaction rate can be estimated.*

If this marching behavior seen from cycle to cycle is caused by a side reaction such as an electrolyte-reduction reaction,

$$S + nLi^{+} + ne^{-} \rightarrow P \qquad 1$$

where *S* represents the solvent (EC, DEC, PC or FEC), *P* represents the reduction product (*e.g.,* $Li_2CO_3$) formed on the active material, and *n* represents the number of electrons transferred in this reaction, the total applied current during the galvanostatic lithiation/delithiation process can be written as:

$$i_{app} = i_{main} + i_{side} \qquad 2$$

where $i_{main}$ and $i_{side}$ correspond to the main-reaction and side-reaction currents, respectively. Similar to the approaches taken by Darling and Newman [37] for the $Li_yMn_2O_4$ system, by Srinivasan *et al.* [38], and by Ta and Newman [39] for the nickel hydroxide system, Tafel kinetics is assumed for the electrolyte-reduction reaction with an equilibrium potential of 0.8 V *vs.* Li/Li$^+$. The current due to this reaction can be written as,

$$i_{side} = -i_{0,side} \exp\left[ -\frac{\alpha_{side} nF}{RT} \left( V - U_{side} \right) \right] \qquad 3$$

In this equation, $i_{0,side}$ represents the side-reaction rate constant, $\alpha_{side}$ is the apparent transfer coefficient for the side reaction, $V$ is the cell potential, and $U_{side}$ is the equilibrium potential for the electrolyte-reduction reaction.

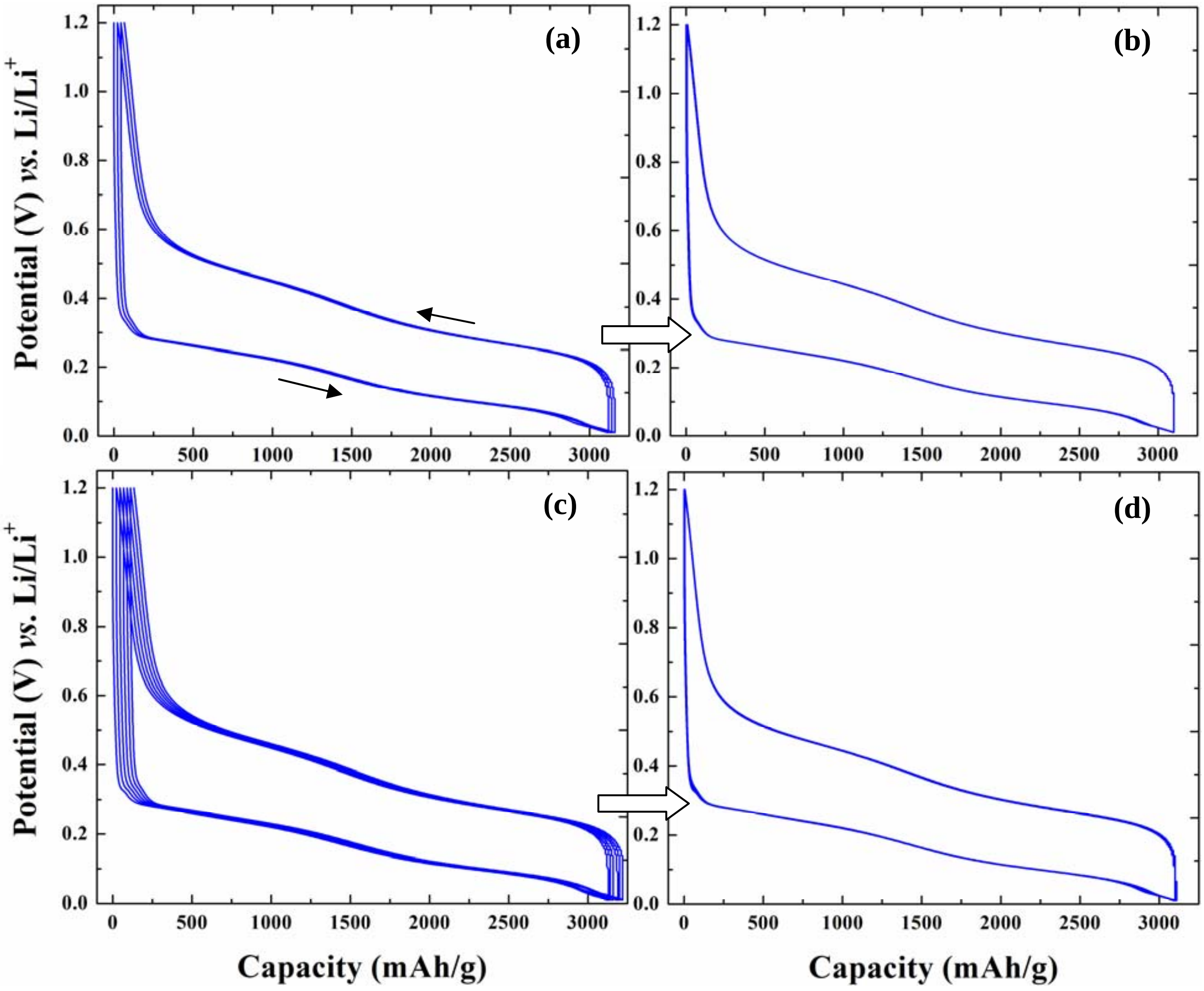

*Figure 4: Cell potential vs. capacity curves for lithiation and delithiation of a well-cycled amorphous silicon thin-film electrode cycled at a C/8 rate between 1.2 and 0.01 V vs. Li/Li$^+$ shown in (a) and (c) are corrected for side reaction using equation 3, and the results are shown in (b) and (d), respectively. The side-reaction correction is done simultaneously using data from three cycles in (a) and five cycles in (c), and the resulting side-reaction rate-constant is ~4.05 x $10^{-13}$ A/cm$^2$ and ~4.25 x $10^{-13}$ A/cm$^2$, respectively. The electrolyte used for this experiment is 1.2 M $LiPF_6$ in EC:DEC (1:2 by % wt.) with 10% FEC. The arrows indicate the cycling direction.*

Though a methodology for estimating $\alpha_{side}$ is described in a separate study [40], a value of 0.5 is assumed for the purposes of this study. An equilibrium potential of 0.8 V *vs.* $Li/Li^+$ for the side reaction was assumed for all three electrolyte formulations because the exact value has not been reported so far. However, in the Tafel regime (*i.e.,* far away from the equilibrium), the equilibrium potential, $U_{side}$, and the side-reaction rate constant, $i_{0,side}$, are not independent of one another [41]. For instance, an assumed value of $U_{side}$ lower than 0.8 V *vs.* $Li/Li^+$ would result in a higher estimate of $i_{0,side}$, and *vice versa*. In view of this, we proceed with $U_{side}$ = 0.8 V *vs.* $Li/Li^+$ and estimate $i_{0,side}$, noting that they are coupled-parameters. Since silicon electrode mostly operates below 0.8 V *vs.* $Li/Li^+$, the side-reaction current is mostly due to a reduction reaction, and because of the existence of potential hysteresis between the lithiation and delithiation processes, coulombic losses due to side-reaction are much higher during lithiation than during delithiation.

This side-reaction current, $i_{side}$, is calculated continuously throughout the cycle (lithiation and delithiation) while assuming an $i_{0,side}$ such that the marching seen in Figure 3a is completely eliminated from the cycling data. Such an attempt is shown in Figure 4a-b. The steady cycling data (for three consecutive cycles) shown in Figure 4a are corrected for the side reaction, and the result is shown in Figure 4b. The corresponding value for $i_{0,side}$ is ~4 x $10^{-13}$ A/cm$^2$. Such an estimation cannot be done on data corresponding to the first few cycles (*i.e.,* cycles 2 to 4) because the delithiation capacity itself is decreasing from cycle to cycle, which is caused by a number of other reasons stated before. Estimation of a side-reaction rate as described above can only be done when there is no cycle-to-cycle variation in delithiation capacity. The accuracy of this estimation procedure can be improved with increasing the number of lithiation/delithiation cycles that are simultaneously used to eliminate the marching behavior (see Figure 4c-d). Furthermore, side-reaction rates estimated using galvanostatic cycling data obtained at different rates between C/8 and 3C were within 5% of one another. The ability of a Tafel equation to completely rectify (or eliminate) the marching behavior seen in this system validates our initial assumption that a reduction reaction causes this behavior.

The limitations of estimating side-reaction currents using this approach must be mentioned. This approach is useful for estimating the charge due to side-reactions only on a *well-cycled* electrode, where the cycle-to-cycle capacity remains constant, and not on a fresh electrode (*i.e.,* during the first formation cycle). The origin of the side-reaction currents on any given cycle in a *well-cycled* electrode is principally due to the instability of the previously formed SEI layer. Because of this, the estimated side-reaction current necessarily reflects both the extent of damage to the previously formed SEI layer (in the form of the newly available surface) as well as electrolyte-reduction kinetics. In order to quantify the latter separately, one need to do precise electrochemical-cycling experiments on a well-defined surface in the neighborhood of the equilibrium potential corresponding to the electrolyte reduction.

Cycling characteristics of two other electrolyte formulations, namely, EC:DEC without the FEC additive, and PC, were studied. The side-reaction current for these two electrolytes were also estimated. A comparison of cycling efficiencies between all three electrolytes used in this study is made in Figure 5. The first-cycle efficiency for the FEC electrolyte is lower than the other two electrolytes indicating a larger charge contribution towards the SEI-layer formation. Subsequently, the FEC electrolyte shows a much higher second-cycle efficiency, whereas, the other two electrolytes exhibit a gradual increase in cycling efficiency over a number of cycles. Side-reaction rates are also lower in FEC compared to the other two electrolytes. Recently, using

X-ray photoelectron spectroscopy (XPS) and time-of-flight secondary-ion mass spectrometry (ToF-SIMS) studies, Nakai *et al.* have reported that the FEC-derived SEI layer consists of lithium fluoride and a polyene compound, and this thin layer protects the Si electrode from oxidation and prevents further decomposition of the electrolyte [35].

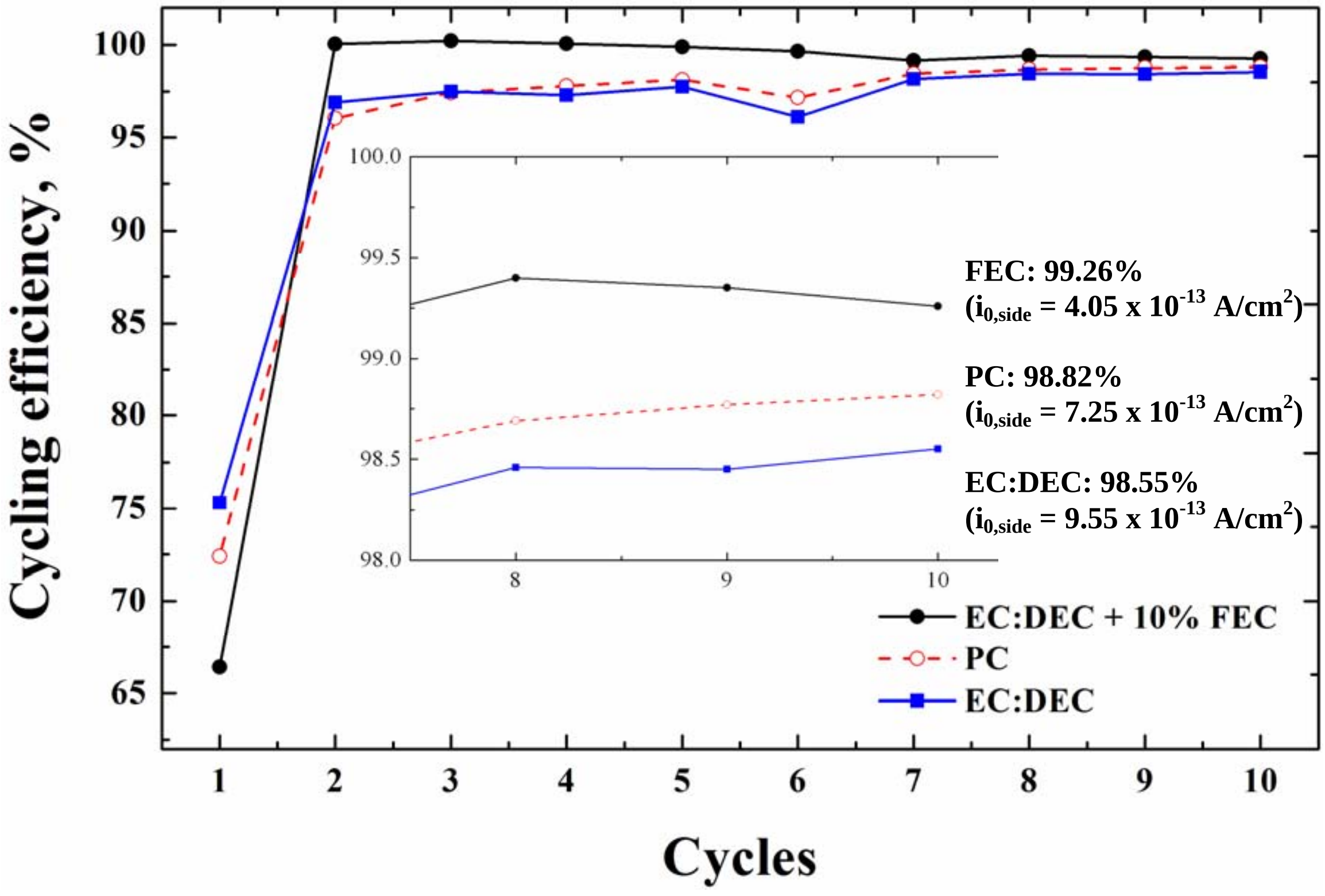

*Figure 5: Cycling efficiency for the first ten cycles is compared for the three different electrolytes studied. Cycling efficiency values for the last three cycles are shown magnified in the inset. Average cycling efficiency, and the estimated side-reaction rate-constants are also shown for the three electrolyte formulations.*

Cycling studies on the FEC electrolyte were done at various temperatures between 5 and 45 °C, and the activation energy for side reaction on a well-cycled Si electrode was estimated to be 20.2±0.5 kcal/mol. The estimate takes into account the error margin in estimating $i_{0,side}$ at the three temperatures. This is comparable to theoretically estimated activation energy reported by Wang *et al.* in similar ethylene-carbonate-based electrolytes [42]. The activation energy was estimated from the Arrhenius plots shown in Figure 6. For comparison, the exchange-current densities corresponding to the lithiation and delithiation reactions estimated using open-circuit-potential-relaxation experiments at 50% SOC are also shown Figure 6. The procedures for estimating the lithiation/delithiation kinetics are beyond the scope of this study and can be found in reference 43. Though a methodology for separating the main and side-reaction currents is provided here, a more detailed study accounting for the concentration dependence of the $i_{0,side}$ term and the precise measurements of $U_{side}$ and $\alpha_{side}$ is currently ongoing will be reported in the future as a separate study.

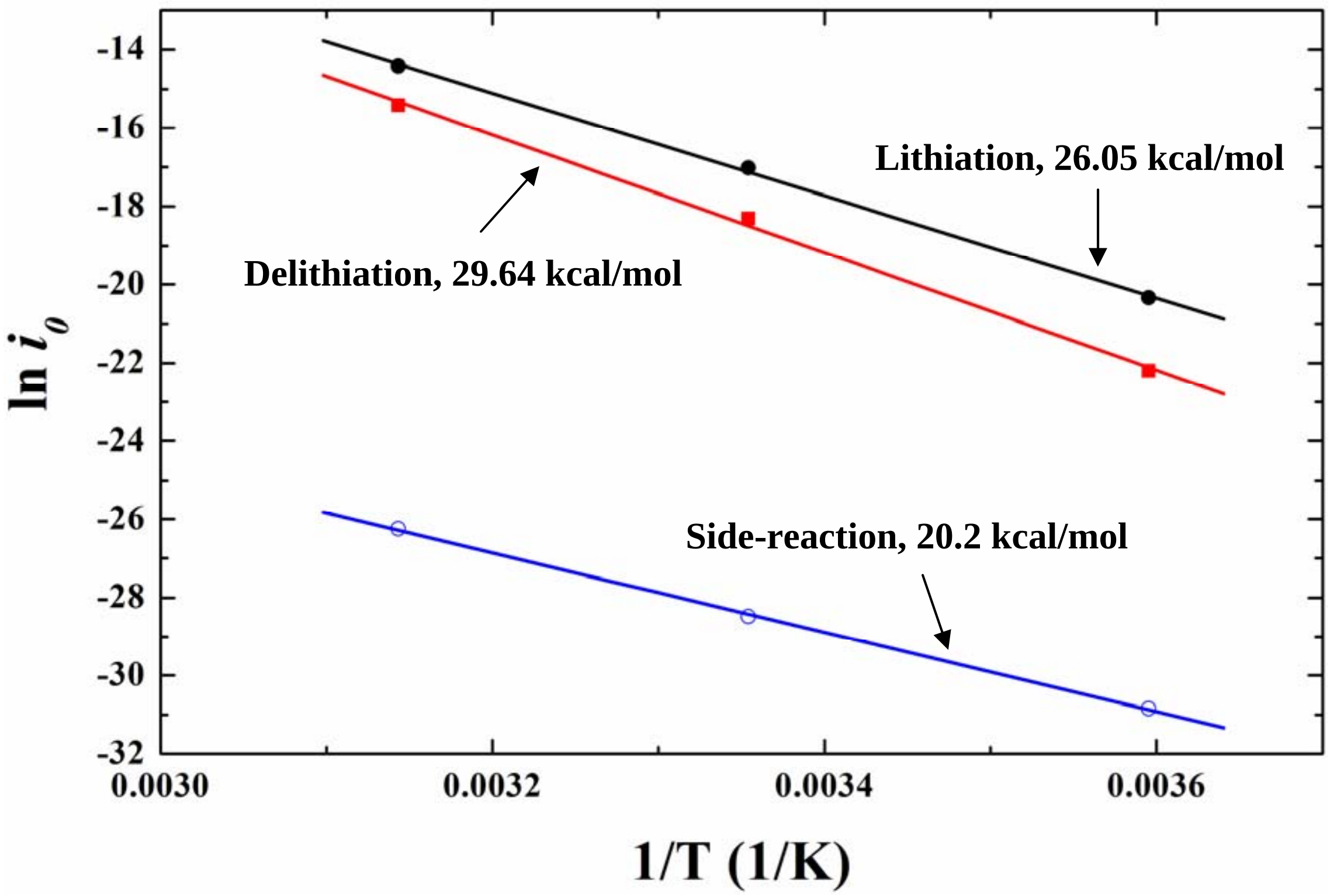

*Figure 6: Arrhenius plot of reaction-rate constant corresponding to the electrolyte reaction (-○-) on, and the lithiation (-●-) and delithiation (-■-) of Si electrode.*

Quantifying side-reaction rates in functioning, well-cycled lithium-ion batteries is important for the following reasons: First, for accurate analysis of cycling data and precise estimation of SOC of the active material, the charge/discharge data need to be accounted for the continuously occurring side-reactions. The procedures described in this article can be used for eliminating contributions due to side reactions from the cycling data. Experimental measurements of side-reaction kinetics also find use in mathematical modeling of capacity fade in lithium-ion batteries [44,45,46], as well as in designing real-time SOC estimation protocols for use in coulomb-counting algorithms and battery-management systems, an important problem for automotive batteries.

Second, the self-discharge (or shelf-life) characteristics of a lithium-ion battery depend on the thermodynamic instability of the active materials as well as on the electrode side-reaction kinetics. The potential of a fully charged silicon anode in a lithium-ion battery is *ca.* 50 mV *vs.* $Li/Li^+$. Since the equilibrium potential for the electrolyte-reduction reaction is much higher at *ca.* 800 mV *vs.* $Li/Li^+$, there is always a driving force for the side-reaction to occur, even in the absence of an external current or load. Electrolytes that aid in the formation of a stable SEI layer, and those that exhibit low side-reaction rates improve the battery's shelf-life characteristics.

Third, side reactions in lithium-battery anodes consume cycleable lithium ions, a limited resource, which when depleted results in capacity fade. Studies on the calendar life of lithium-ion batteries with graphite anodes suggest that side-reactions play a significant role in the loss of reversible capacity [47,48]. It is expected that side reactions will play an even more detrimental role in silicon electrodes which routinely undergo much larger volume expansion (up to 270% *vs.* 10% for graphite) and experience large stresses [24,30]. Huge volume changes cause the stretching of surface SEI films (usually comprised of organic and inorganic lithium salts), that are brittle with low flexibility, resulting in their breaking. As a consequence, underlying fresh surfaces are exposed to the electrolyte resulting in more SEI formation, which increases the impedance of the electrode. In view of this, it is important to identify and evaluate surface coatings capable of withstanding repeated volume expansion/contraction cycles, to enable better cycling characteristics and improve battery life without sacrificing performance.

## 4. Conclusions

Continuously occurring side reactions account for the reduction in cycling efficiency in silicon anodes in lithium-ion batteries. A methodology for estimating the side-reaction current on a functioning, well-cycled silicon electrode in a Si/Li half-cell is presented. The rate of this side reaction is estimated for three common electrolyte formulations; the electrolyte containing EC:DEC (3:7 wt%) with 10% FEC additive is shown to exhibit the least side-reaction current and consequently, the best cycling efficiency. The importance of quantifying side-reaction current in silicon anodes is discussed in the context of self-discharge, capacity fade, designing SOC indicator algorithms, and precise mathematical modeling of lithium-ion batteries.

## 5. Acknowledgements

Financial support in the form of a postdoctoral fellowship from the Assistant Secretary for Energy Efficiency and Renewable Energy, Office of Vehicle Technologies, the United States Department of Energy under contract no. DE-AC02-05CH11231 through V. Srinivasan is gratefully acknowledged. Helpful discussions with P.R. Guduru (Brown University) and V. Srinivasan (Lawrence Berkeley National Laboratory) are gratefully acknowledged. Silicon thin-films were fabricated at the Micro-fabrication Laboratory, Department of Electrical Engineering, University of California, Berkeley. Scanning electron micrographs were obtained at the National Center for Electron Microscopy, Lawrence Berkeley National Laboratory.